\documentclass[twocolumn,preprintnumbers,amsmath,amssymb]{revtex4-2}                                                                                                                                                                                                                                                                                                                                                                                                                                                                                                                                                                                                                                                                                                                                                                                                                                                                                                                                                                                             %\documentclass[manuscript]{revtex4-2} 

\usepackage{graphicx}  
\usepackage{sidecap}
\begin{document}
%\emph{•}
%   
\title{Hot-electron resonant  terahertz bolometric detection in the graphene/black-AsP  field-effect transistors with a floating gate
}
\author{V.~Ryzhii$^{1}$, C.~Tang$^1$, T.~Otsuji$^1$, M.~Ryzhii$^{2}$, 
V.~Mitin$^3$,  and M.~S.~Shur$^4$}
\address{
$^1$Research Institute of Electrical Communication,~Tohoku University,~Sendai~ 980-8577, 
Japan\\
$^2$Department of Computer Science and Engineering, University of Aizu, Aizu-Wakamatsu 965-8580, Japan\\
$^3$Department of Electrical Engineering, University at Buffalo, SUNY, Buffalo, New York 14260 USA\\
$^4$Department of Electrical,~Computer,~and~Systems~Engineering, Rensselaer Polytechnic Institute,~Troy,~New York~12180,~USA
}

\begin{abstract} 
\normalsize We evaluate the terahertz (THz) detectors based on 
field effect transistor (FET) with the graphene channel {GC} and a floating
metal gate (MG) separated from  the GC by a black-phosphorus  (b-P) or black-arsenic (b-As) barrier layer (BL).
The operation of these GC-FETs is associated with the heating of the two- dimensional electron gas in the
GC by impinging THz radiation leading to thermionic emission of the hot electrons from the GC  to the MG. This results in the variation of the floating gate potential, which affects the source-drain current.
At the THz radiation frequencies close to the plasmonic resonance frequencies in the gated GC, the variation of the source-drain current and, hence, the detector responsivity can be resonantly large. 
\end{abstract} 
\maketitle
%\newpage\newpage

\begin{figure}[t]\centering
\includegraphics[width=6.9cm]{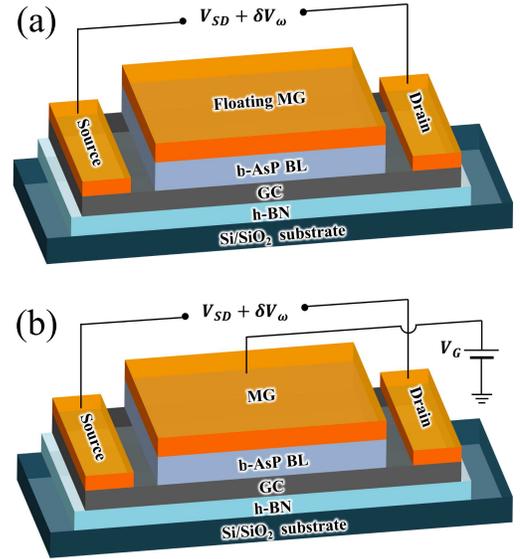}
\caption{
 Schematic view of  the GC/b-AsP FET detectors (a) with the floating  MG and (b) with the MG biased by the gate voltage $V_G$~\cite{16}.
} 
\label{F1}
\end{figure}

\begin{figure}\centering
\includegraphics[width=6.9cm]{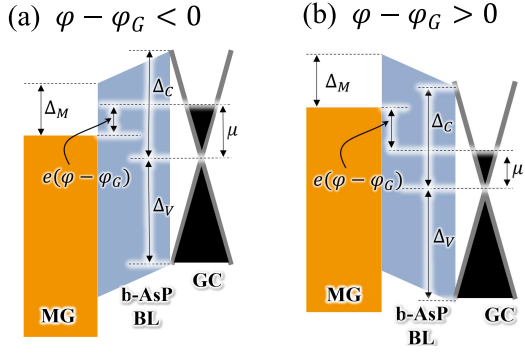}
\vspace{+5mm}
\caption{
 Band diagrams
of GC/b-AsP FET detector with the floating MG [shown in Fig.~1(a)] near (a) the source  ($x \gtrsim -L$, the GC potential $\varphi - \varphi_G< 0$)and (b) the drain ($x \lesssim L$, the GC potential $\varphi  -\varphi_G > 0$ ).
} 
\label{F2}
\end{figure}

\section{Introduction}
The specific properties of graphene channel (GCs)~\cite{1,2,3} and   black-P  (b-P), black-As (b-As) , or black-AsP (b-AsP) layers~\cite{4,5,6,7,8} open up prospects for  devices based on the GCs~(see, for example, the review~\cite{9}) and 
 on the GC/b-AsP  heterostructures~\cite{6}, including the   electron devices  using
the real-space transfer over the b-AsP layers~\cite{10,11} 
and different optoelectronic devices~\cite{12,13,14,15}. 
Due to relatively low energy barriers for the electrons and holes at
the  GC/b-AsP interface,
the thermionic emission through such an interface can be effective, particularly,
enabling the creation of the GC/b-AsP- bolometric terahertz (THz) detectors.\\
In this paper, we evaluate the characteristics of the bolometric detectors 
based on the field-effect transistor (FET) structures with the GC, b-AsP
barrier layer (BL), and floating metal gate (MG).
Similar GC/b-AsP FET detectors were recently proposed and analyzed by us~\cite{16}.
The principal  difference between the GC/b-AsP FETs considered previously and 
the GC/b-AsP FETs under consideration here is the floating MG.
%%%%
The idea of using MG in graphene bolometers has been applied to pyroelectric graphene mid-infrared
detectors. In these detectors, pyroelectric substrate charge is collected by a floating gate~\cite{17}.

In contrast, the operation of the  bolometric detectors considered in this paper 
 is associated with the thermionic emission of the electrons heated by the impinging THz radiation from
the GC into the MG via the b-AsP BL. 
However, contrary to the devices studied in~\cite{16}, in which the gate current serves as the detected signal,
in the detector with the floating MG considered here the detected signal is associated with the variations of the source-drain current in the GC stimulated by the varying potential of the MG. The potential of the latter is controlled by the thermionic emission from the GC, which reinforces with increasing THz power. 
This effect can become rather strong at the plasmonic oscillations resonantly  excited by the impinging THz radiation in the gated GC~\cite{18,19,20,21,22}.
The features of the  GC-FET detector operation with the floating gate require the development of a fairly different device model.
Using this model, we calculate  the signal current and the detector responsivity as functions of the structural parameters.  
As demonstrated, the floating gate GC/b-AsP FET detectors might exhibit elevated
values of the responsivity, particularly, at the plasmonic resonances.
We also compare the performance of GC/b-AsP FET detectors  with the floating and biased MGs.

\section{Electron transport}

We consider the GC/b-AsP FETs with the floating MG and the b-AsP gate BL.
Figures~1 and 2 schematically show
 the cross-section of the device structures and their  band diagrams.
The bias voltage $V_{SD}$ and the signal voltage $\delta V_{\omega}$ are applied
between the FET source and drain as shown in Fig.~1(a).
The signal voltage is produced by an antenna receiving the impinging THz radiation with the frequency $\omega$.
The GC of the FETs is doped by donors.
For definiteness, the work functions of the  gate metal and the b-AsP BL,   and the GC doping level (the electron Fermi energy $\mu_D$ in the GC in equilibrium when no bias is applied) are chosen to provide the band alignment in the equilibrium. This corresponds to $\Delta_M =\Delta_C - \mu_D$,
where $\Delta_M$ and $\Delta_C$ are the differences between the work functions of the gate metal and the  b-AsP BL and between the b-AsP BL and the GC.

At the source-drain bias voltage $V_{SD}$ and the  THz irradiation,
the source-drain current, $J_{SD}$, and  the electron effective  temperature, $T$, in the GC averaged over the THz radiation period $2\pi/\omega$ can be presented as

\begin{eqnarray}\label{eq1}
J_{SD} = {\overline J}_{SD} + \langle\delta J_{\omega}\rangle, \qquad T = {\overline T} + \langle\delta T_{\omega}\rangle.
\end{eqnarray} 
Here 
${\overline J}_{SD} = J_0 +  \Delta{\overline J}_{DC}$ and 
${\overline T} = T_0 + \Delta{\overline T}$, $J_0$ is the source-drain current at the 2DEG effective temperature equal to the lattice
 $T_0$,
 $\Delta{\overline J}_{DC}$ and $\Delta{\overline T}$ are the pertinent
 current and temperature variations, and
 $\langle\delta J_{SD}\rangle$ and  $ \langle\delta T_{\omega}\rangle$
 are the variations  caused by the source-drain bias voltage $V_{SD}$ and the signal voltage
 $\delta V_{\omega}$.

 The source-drain current $ J_{SD}$ per unit of the GC width 
is governed by the following equations:
 
\begin{eqnarray}\label{eq2}
\frac{d J_{SD}}{dx} = - j, \qquad J_{SD} = -  \sigma \frac{d\varphi}{dx},
\end{eqnarray} 
where 
$j$ is the density of the thermionic current between the GC and the MG,
$\sigma =\sigma_D(\mu/\mu_D)$ and  
$\sigma_D =(e^2 \mu_D/\pi\hbar^2\nu)$ 
 is the electron Drude conductivity in equilibrium  with $\nu$
 being the characteristic electron scattering frequency in the GC (the inverse electron momentum relaxation time), and $\mu$ is the electron Fermi energy, which generally differs from $\mu_D$ due to the MG charging.
 
The averaged GC potential $\varphi$ (dependent on the coordinate $x$ directed along the GC) satisfies the following  conditions at the source and drain contacts:

\begin{eqnarray}\label{eq3}
\varphi|_{x =\pm L}  
= \pm 
\frac{V_{SD}}{2},
\end{eqnarray} 
where $2L$ is the spacing between the contacts (the GC  length).
Considering  the difference between the GC potential $\varphi$
and the MG potential $\varphi_G$ and accounting for 
 the quantum capacitance~\cite{23,24} of the gated GC, 
at not too-large  potential swing $(\varphi-\varphi_G)$
we obtain

\begin{eqnarray}\label{eq4}
\mu \simeq \mu_D - {\varkappa}\,e(\varphi - \varphi_G). 
\end{eqnarray} 
Here ${\varkappa}= \mu_0/(\mu_0+\mu_D)  $, $\mu_0 = \kappa\,\hbar^2v_W^2/8e^2W$, $\kappa$ is the BL dielectric constant, and $v_W \simeq 10^8$~cm/s
 is the electron velocity in GCs. This implies that an increase in $\varphi$ leads to an increase in the 2DEG density and, hence, its Fermi energy $\mu$. The contribution  of the quantum capacitance to Eq.~(4) is characterized by a factor $\mu_0/\mu_D \propto W^{-1}$.

Since the MG is disconnected (floating MG),

\begin{eqnarray}\label{eq5}
\int_{-L}^{L}dx j = 0.
\end{eqnarray}

Due to the trapezoid shape of the barrier between the GC and the MG,
 the potential barrier heights for the electron emitted from the GC
 and  the MG,  $\Delta_{BL}^{\leftarrow}$ and
$\Delta_{BL}^{\rightarrow}$, are equal to: 
 
 $\Delta_{BL}^{\leftarrow} = \Delta_M +e(\varphi -\varphi_G) $ and
$\Delta_{BL}^{\rightarrow} = \Delta_M$
for $\varphi > \varphi_G$,
 and 
 
 $\Delta_{BL}^{\leftarrow}  =\Delta_M - (\mu - \mu_D)$ and $\Delta_{BL}^{\rightarrow} =  \Delta_M - (\mu - \mu_D) - e(\varphi - \varphi_G)$
for $\varphi < \varphi_G$.

In this situation,
the density of the thermionic current, $j$, between the GC and the MG is given by

\begin{eqnarray}\label{eq6}
j = j^{m}\biggl[\exp\biggl(-\frac{\Delta_M +e(\varphi - \varphi_G)}{T}\biggr)\nonumber\\
-
\exp\biggl(-\frac{\Delta_M}{T_0}\biggr)\biggr]
\end{eqnarray} 
when $\varphi - \varphi_G >0$, and

\begin{eqnarray}\label{eq7}
j = j^{m}\biggl[\exp\biggl(-\frac{\Delta_M - (\mu -\mu_D) }{T}\biggr)
\nonumber\\
-
\exp\biggl(-\frac{\Delta_M - (\mu - \mu_D) - e(\varphi- \varphi_G)}{T_0}\biggr)\biggr]
\end{eqnarray} 
when $\varphi - \varphi_G <0$.
Here $j^{m} \simeq e\Sigma/\tau_{\bot}$ is the maximum current density,
$\Sigma$ is the electron density in the G-channel, and $\tau_{\bot}$ is the characteristic try-to-escape time from the G-channel.
 From Eqs.~(2), (4), (6), and (7), we obtain
 
\begin{eqnarray}\label{eq8}
-\frac{L^2}{\mu_D}\frac{d}{dx}
\biggl[\biggl(1 -
{\varkappa}\,\frac{e(\varphi -\varphi_G)}{\mu_D}\biggr)\frac{d\varphi}{dx}\biggr]\nonumber\\
 =  \eta
\biggl[
\exp\biggl(\frac{\Delta_M(T-T_0)}{T_0 T}\biggr)
\exp\biggl(-\frac{e(\varphi - \varphi_G)}{ T}\biggr) -1\biggr]
%\eta
%\biggl[\exp\biggl(-\frac{\Delta_M + \mu_D{\overline \psi}}{{\overline T}}\biggr)
%-
%\exp\biggl(-\frac{\Delta_M}{T_0}\biggr)\biggr]
\end{eqnarray}
when $\varphi - \varphi_G >0$, and

\begin{eqnarray}\label{eq9}
-\frac{L^2}{\mu_D}\frac{d}{dx}
\biggl[\biggl(1 -
{\varkappa}\,\frac{e(\varphi -\varphi_G)}{\mu_D}\biggr)\frac{d\varphi}{dx}\biggr]\nonumber\\
 =
\eta\biggl[
\exp\biggl(\frac{\Delta_M (T-T_0)}{T_0T}\biggr)
\exp\biggl(-{\varkappa}\frac{e(\varphi - \varphi_G}{T}\biggr)
\nonumber\\
 -\exp\biggl((1-{\varkappa})\frac{e(\varphi - \varphi_G)}{T_0}\biggr)\biggr] 
\end{eqnarray} 
when $\varphi - \varphi_G <0$. 

 Here
\begin{eqnarray}\label{eq10}
\eta = \frac{ej^{m}}{\mu_D \sigma_D}\exp\biggl(-\frac{\Delta_M}{T_0}\biggr)=\frac{\nu\,L^2}{\tau_{\bot}v_W^2}\exp\biggl(-\frac{\Delta_M}{T_0}\biggr).
\end{eqnarray} 
Setting, for example, $\Delta_M = 85$~meV, $T_0 = 25$~meV, $\nu = 1$~ps$^{-1}$, $\tau_{\bot} = 10$~ps, $L = 1.0~\mu$m, 
$\kappa = 4-6$,  $W = 10$~nm, and $\mu_D = 140$~meV,
we obtain
$\eta \simeq (3.3)\times 10^{-3}$ and 
${\varkappa} \simeq 0.088 - 0.127$ [$\mu_0 \simeq (13.6 - 20.4)$~meV].

At low or moderate bias source-drain voltages and THz radiation powers, $ \psi$ and $| T - T_0|/ T_0$ are small. In this case, linearizing Eqs.~(8) and (9), we arrive at

\begin{eqnarray}\label{eq11}
L^2\frac{d^2}{dx^2}\biggl[ e(\varphi -\varphi_G) +\frac{\varkappa}{2}\frac{e^2(\varphi -\varphi_G)^2}{\mu_D}\biggr]\nonumber\\
 \simeq  
\frac{\eta\mu_D}{T_0}
\biggl[\frac{\Delta_M}{T_0}( T - T_0)
- e(\varphi -\varphi_G)
\biggr]
%\nonumber\\ - \frac{L^2{\varkappa}}{2}\frac{d^2{\overline \psi}_0^2}{dx^2},
%\qquad
\end{eqnarray}
Equation~(11) corresponds to the thermionic current density

\begin{eqnarray}\label{eq12}
j \simeq \frac{j^m}{T_0} \biggl[\frac{\Delta_M}{T_0}( T - T_0)
- e(\varphi -\varphi_G)
\biggr].
\end{eqnarray}
Using Eq.~(11) with Eq.~(3) and taking into account the smallness of parameter $\eta$ [i.e., neglecting the term in the right-hand side of Eq.~(11)],  for the source-drain current $J_{SD} = \sigma (HV_{SD}/2L) = \sigma_D(\mu/\mu_D)(HV_{SD}/2L)$ with the pertinent  accuracy we obtain

\begin{eqnarray}\label{eq13}
 J_{SD}\simeq %\sigma_D(1 + \varkappa\psi_G)\frac{V_{SD}}{2L}.
\sigma_D\biggl(1 + \varkappa \frac{e\varphi_G}{\mu_D}\biggr)\frac{H}{2L}V_{SD}.
\end{eqnarray}
Using Eqs.~(5) and (12), we find the MG potential: 

\begin{eqnarray}\label{eq14}
e\varphi_G = \frac{eV_{SD}}{4} -\frac{\Delta_M}{2L}\int_{-L}^Ldx\frac{(T-T_0)}{T_0}.
\end{eqnarray}
The latter equation corresponds to an increase in the source-drain current with increasing gate potential $\varphi_G$ (due to an increase in the Fermi energy and, hence, the G-channel conductivity). One can see that an increase in the 2DEG
effective temperature leads to the intensification of the electron transfer from the G-channel to the gate which results in its lower potential.

Equations~(13) and (14) for the  source-drain current components
  yield

\begin{eqnarray}\label{eq15}
{\overline  J}_{SD} \simeq J_0\biggl[1 + \frac{\varkappa}{\mu_D}\biggl(eV_{SD} -\frac{\Delta_M}{2L}\int_{-L}^Ldx\frac{ ({\overline T} - T_0}{T_0}\biggr)\biggr]
\end{eqnarray}
and

\begin{eqnarray}\label{eq16}
\langle\delta J_{\omega}\rangle \simeq - J_0\frac{\varkappa\Delta_M}{\mu_D}\frac{\langle\langle \delta T_{\omega}\rangle\rangle}{T_0}
%\biggl(\frac{1}{2L}\int_{-L}^Ldx\frac{\langle\delta T_{\omega}\rangle}{T_0}\biggr) 
\end{eqnarray}
with $J_{0} =\sigma_DV_{SD}(H/2L)$ and $\langle\langle \delta T_{\omega}\rangle\rangle = \int_{-L}^{L}dx\langle \delta T_{\omega}\rangle/2L$ and $H$  being the 
effective temperature average of the THz period and the GC length and  the GC width, respectively. The quantity $\langle\delta J_{\omega}\rangle$ given by Eq.~(16) represents the response of the GC-FET to
the impinging THz radiation.

\begin{table*}[]
\centering
\caption{\label{table} Parameters of the GC/b-As and GC/b-P FET detectors and their responsivities.}
\begin{tabular}{|r|c|c|c|c|c|c|c|c|c|c|c|}
\hline
Structure\,&   $\Delta_M$(meV)& $\Delta_C$(meV)& $\mu_D$~(meV)& $L$ ($\mu$m)  &$\Omega/2\pi$ (THz)& $\nu$ (ps$^{-1}$)& $\theta$ &
${\mathcal L}$ ($\mu$m) &$R_{\Omega}^V$ (V/W)\\ 
\hline
GC/b-P/Al\,&	85&	225& 140 & 1.0& 1.136& 1.0 - 2.5 &1.02 &1.58 - 1.00&$(2.1 - 1.8) \times10^3$\\ 
\hline
GC/b-As/Ti& 70&  190& 120& 1.0 &1.052&1.0 -2.5 &1.29 &0.93 - 0.59&$(2.7 - 2.3)\times 10^3$\\
\hline
\end{tabular}
\end{table*}

\section{Electron heating and heat transport}
The electron heat transport  equation can be presented as

\begin{eqnarray}\label{eq17}
-h\frac{d^2T}{dx^2} 
+\frac{ T-T_0}{\tau_{\varepsilon}}
\nonumber\\
+
\frac{\Delta_C\Delta_M}{\tau_{\bot}T_0}
\exp\biggl(-\frac{\Delta_M}{T_0}\biggr)
\biggl[\frac{\Delta_M}{T_0}( T - T_0)
- e(\varphi -\varphi_G)
\biggr]
\nonumber\\
 \simeq
 \frac{\sigma}{\Sigma}\biggl[\biggl(\frac{V_{SD}}{2L}\biggr)^2 + \frac{{\rm Re} \sigma_{\omega}}{\sigma}\langle|\delta E_{\omega}|^2\rangle\biggr].
\end{eqnarray}
For the variation of the  2DEG averaged effective temperature $\langle \delta T_{\omega}\rangle$,  in view of  Eqs.~(5) and (12), Eq.~(17) yields

\begin{eqnarray}\label{eq18}
-h\frac{d^2\langle \delta T_{\omega}\rangle}{dx^2} 
+\frac{\langle \delta T_{\omega}\rangle}{\tau_{\varepsilon}}
\nonumber\\
+
\frac{\Delta_C\Delta_M}{\tau_{\bot}T_0}
\exp\biggl(-\frac{\Delta_M}{T_0}\biggr)\biggl(\frac{\langle \delta T_{\omega}\rangle -\langle\langle \delta T_{\omega}\rangle\rangle}{T_0}
\biggr)
\nonumber\\
 \simeq
 \frac{{\rm Re }\sigma_{\omega}}{\Sigma}\langle|\delta E_{\omega}|^2\rangle).
\end{eqnarray}
Here $h \simeq v_W^2/2\nu$ is the electron thermal conductivity (per electron), $\tau_{\varepsilon}$ is the electron energy relaxation time, Re$\sigma_{\omega} =\sigma_D\,\nu^2/(\nu^2+\omega^2)$ is the real part of the 2DEG ac conductivity,  and
 $\delta E_{\omega}$ is the signal electric 
fields in the G-channel created due to  the THz signals.
The first, second, and third  terms in the left-hand side of Eq.~(18)
are associated with the removal of the electron heat through the contact
(due to a substantial electron lateral heat conductivity along the GC~\cite{25,26}), the 
transfer to the lattice (primarily due to the interaction with optical phonons in the GC and the interface optical phonons~\cite{27,28,29,30,31}) and   the MG over the BL (i.e., corresponding to the Peltier cooling~\cite{32,33}),  respectively.
The term on the right-hand side of Eq.~(13) corresponds to the 2DEG Joule heating.

We use the following  boundary conditions for Eq.~(18): 

\begin{eqnarray}\label{eq19}
\langle \delta T_{\omega}\rangle|_{x= \pm L} = 0.
\end{eqnarray}

For the THz radiation asymmetric input via the antenna corresponding to the signal potential at the contacts equal to $\pm\delta V_{\omega}/2$, 
accounting for the excitation of plasmonic oscillations in the GC  we obtain

\begin{eqnarray}\label{eq20}
 \langle |\delta E_{\omega}|^2\rangle =\frac{1}{2}\biggl(\frac{\delta V_{\omega}}{2L}\biggr)^2\biggl|
\frac{\gamma_{\omega}\cos(\gamma_{\omega}x/L)}{\sin\gamma_{\omega}}\biggr|^2.
\end{eqnarray}
Here
 $\gamma_{\omega} =\pi\sqrt{\omega(\omega+i\nu)}/\Omega$ and
$\Omega=(2\pi\,e/\hbar\,L)\sqrt{\mu\,W/\kappa}$ are the effective wavenumber and the plasmonic frequency, respectively, with
$\kappa$ and $W$ being the dielectric constant of the BL  and its thickness,

Restricting our consideration by the most interesting case of the pronounced fundamental plasmonic resonance in the G-channel ($\omega = \Omega \gg \nu$) and using Eq~(15), we obtain

\begin{eqnarray}\label{eq21}
{\rm Re}\sigma_{\Omega} \langle |\delta E_{\Omega}|^2\rangle \simeq
2\sigma_D\cos^2\biggl(\frac{\pi\,x}{L}\biggr)\biggl(\frac{\delta V_{\Omega}}{2L}\biggr)^2.
\end{eqnarray}
One needs to note that the Joule power at the plasmonic resonance given
by Eq.~(21)  exceeds that at low frequencies (at least near $x \simeq 0$) by a factor $\sim(2\Omega/\pi\nu)^2$. 

Solving Eq.~(18) accounting for boundary condition~(19) and Eq.~(21),
for the values $\langle\langle \delta T_{\omega}\rangle\rangle$ at the fundamental plasmonic resonance we obtain

\begin{eqnarray}\label{eq22}
\langle\langle \delta T_{\Omega}\rangle\rangle 
\simeq 
\frac{2\pi\sigma_D\hbar^2v_W^2}{\mu_D^2}\frac{\tau_{\varepsilon}\,\Theta}{(1+\theta)}
\biggl(\frac{\delta V_{\Omega}}{2L}\biggr)^2.
\end{eqnarray}
Here
\begin{eqnarray}\label{eq23}
\Theta = \biggl[\frac{1 - \displaystyle\frac{{\mathcal L}}{L} \tanh\biggl(\frac{L}
{\mathcal L}\biggr)}{1 + \displaystyle\frac{{\mathcal L}}{L} \tanh\biggl(\frac{L}{\mathcal L}\biggr)}\biggr]
\end{eqnarray}
is the factor characterizing the role of electron thermal transport.
and
$$
 \theta =  \frac{\tau_{\varepsilon}}{\tau_{\bot}} \frac{\Delta_C\Delta_M}{T_0^2}\exp\biggl(-\frac{\Delta_M}{T_0}\biggr),
  \qquad {\mathcal L} = \sqrt{\frac{h\tau_{\varepsilon}}{(1 +\theta)}},
$$
The characteristic length ${\mathcal L}$ is the electron heat transfer (cooling) length. 

Equations~(16) and (22) yield

\begin{eqnarray}\label{eq24}
-\langle \delta J_{\Omega}\rangle \simeq \frac{2\pi\sigma_D^2\hbar^2v_W^2}{\mu_D^2T_0}\biggl(\frac{\varkappa\Delta_M}{\mu_D}\biggr)
%\nonumber\\
%
\frac{\tau_{\varepsilon}\,\Theta}{(1+\theta)}
\biggl(\frac{HV_{SD}}{2L}\biggr)
\biggl(\frac{\delta V_{\Omega}}{2L}\biggr)^2.\qquad
\end{eqnarray}
The sign "minus" in Eq.~(24)  reflects the fact that the THz irradiation leads to an increase in the electron effective temperature, reinforcement
of the electron emission from the GC and, hence, to a negative charging of the MG. The latter, in turn, decreases the source-drain current in the donor-doped GC.

\section{Detector responsivity}

Considering that for the half-wavelength dipole antenna  with the gain $g$ one obtains
$\delta V_{\Omega}^2 =32 P_{\Omega}/gc$,
where $P_{\Omega}$ is the THz power at the frequency $\omega = \Omega$ collected by  the detector antenna 
and $c$ is the speed of light in vacuum, and accounting for that
the GC channel resistance is equal to $r_{SD} = 2L/H\,\sigma_D$, 
for the detector voltage responsivity $R_{\Omega}^V = |\langle \delta J_{\Omega}\rangle|\,r_{SD}/P_{\Omega}$  (at the radiation frequency corresponding to the fundamental plasmonic resonance), we obtain

\begin{eqnarray}\label{eq25}
R_{\Omega}^V\simeq\frac{16e^2}{gc\mu_D^2}
\biggl(\frac{L_{\varepsilon}}{L}\biggr)^2
\biggl(\frac{\varkappa\Delta_M}{T_0}\biggr)
\,\Theta\,
V_{SD}\nonumber\\
=
\frac{16}{137g}\frac{\hbar}{\mu_D^2}
\biggl(\frac{L_{\varepsilon}}{L}\biggr)^2
\biggl(\frac{\varkappa\Delta_M}{T_0}\biggr)\,
\Theta\,
V_{SD}.
\end{eqnarray}
Here $L_{\varepsilon} =\displaystyle \sqrt{\frac{v_W^2\tau_{\varepsilon}}{\nu(1+\theta)}}$.

The quantities $\Delta_M$, $\mu_D$, and $\varkappa$, are determined by the material of the MG and 
the  molar fractions of As in the BL (due to the condition $\Delta_M = \Delta_C - \mu_D$ assumed in our model).

Examples of the parameters of 
the GC-FET detectors based on Al/b-P/GC  and Ti/b-As/GC heterostructures (see, for example, Refs.~[34] and [35]) and the estimates of their resonant responsivity are listed in Table I.
We assume also  
 that $\kappa = 4$,
$\tau_{\varepsilon} = 10$~ps,
$\tau_{\bot} = 10$~ps, $W = 10$~nm, $T_0 =25$~meV ($\sim 300$~K),  and $V_{SD} = 1.6$~V.
The above  parameters (with $\nu = 1$~ps$^{-1}$ and  $H = 2L$ or with $\nu = 2.5$~ps$^{-1}$ and $H = 5L$) correspond to  the GC-FET detector resistances $r_{SD} \simeq 55~\Omega$ and $r_{SD}\simeq 64~\Omega$, respectively (at $H = 2L$). 
One needs to note that at the above parameters the  electron thermal transport factor
depending on the ratio ${\mathcal L}/L$ in Eq.~(23) is rather small (about  0.061 -- 0.151 at $\nu = 1$~ps$^{-1}$) substantially decreasing the responsivity. The role of the
electron cooling due to the thermal transport to the side contacts can be decreased by increasing $\nu$ (this decreases the electron thermal conductivity) or choosing the longer GC length~$L$.

Figure~3 shows the responsivity of the GC/b-AsP and GC/b-As FETs  
with a floating MG  at $\omega = \Omega$ (i.e., at the fundamental plasmonic resonance) calculated for the main parameters corresponding to Table I but  for different electron scattering frequencies $\nu$. For the definiteness,
we set $V_{SD} = 1.6$~V.
 The inset in Fig.~3 shows how the role  of the electron thermal transport along the GC changes with varying scattering frequencies.
The pertinent dependence is associated with the $h$ vs $\nu$ dependence.
As follows from Fig.~3, an increase in $\Theta$ (i.e., a weakening of the electron heat transfer to the source and drain contacts when $\nu$ becomes larger) leads to slowing 
of the $R_{\Omega}^V$ vs $\nu$ dependence. 
%%%%%%%%!
Thus, a relatively weak dependence of $R_{\Omega}^V$ on $\nu$ is interpreted by   the decrease in  the electron system Joule heating in the GC
by the signal electric field (because of $\sigma_D \propto \nu^{-1}$)
accompanied with a decrease in the power transferred to the source and drain contacts.

\begin{figure}\centering
\includegraphics[width=8.0cm]{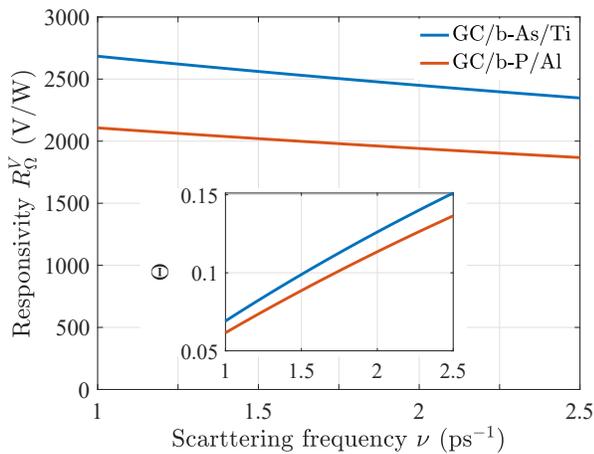}
\caption{Resonant responsivity $R_{\Omega}$  of the GC/b-As/Ti (blue line) and GC/b-P/Al (red line) FET detectors and  electron thermal transport factor $\Theta$ (inset)
 as  functions of  electron scattering frequency $\nu$.} 
\label{F3}
\end{figure}

\begin{figure}\centering
\includegraphics[width=8.0cm]{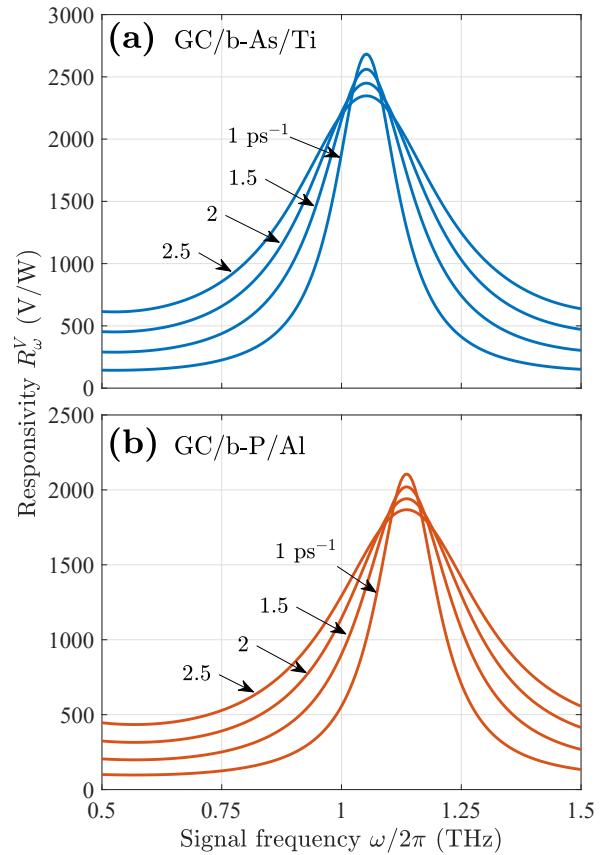}
\caption{
Responsivity $R_{\omega}$  of the  (a) GC/b-As/Ti  and (b) GC/b-P/Al FET detectors vs signal frequency $\omega/2\pi$ for different values of  electron scattering frequency $\nu$.
} 
\label{F4}
\end{figure}

Figure~4 shows the spectral dependences of the   responsivity, 
$R_{\omega}^V$, of the  GC/b-As and GC/b-P detectors calculated for different $\nu$ and the same parameters as for Fig.~3. We limited our consideration by the signal frequencies around the fundamental plasmonic resonance, where  the obtained  dependences exhibit pronounced maxima provided
that $\Omega \gg \nu$. As seen from Fig.~4, an increase in $\nu$ gives rise to a smearing of the resonant peak. 
In a wider frequency range, the responsivity of the detectors under consideration
 is a profoundly oscillatory  function of the radiation signal frequency $\omega$ with a set of the  maxima at the plasmonic resonances $\omega = n\Omega$ ($n$ is the resonance index). These oscillations are  described by the  relation, which follows from the above equations:

\begin{eqnarray}\label{eq26}
R_{\omega}^V \propto {\rm Re}\sigma_{\omega}\biggl|\frac{\gamma_{\omega}\cos(\gamma_{\omega}x/L)}{\sin\gamma_{\omega}}\biggr|^2.
\end{eqnarray}
As follows from Eq.~(26),
the dependences $R_{\omega}^V$ vs $\omega/2\pi$
exhibit the alternation of sharp maxima and relatively deep minima.
At the intermediate  frequencies, the responsivity at the  minima,  is smaller than the resonant responsivity by a factor of $(\pi\nu/2\Omega)^2 \ll 1$.
%%%%%%%!!!!
At elevated collision frequencies $\nu$, the spectral characteristics of $R_{\omega}^V$  become smoother. 
However, up to $(\pi\nu/2\Omega)^2 \sim 1$ (this corresponds to $\nu \sim 4$~ps$^{-1}$,  at the plasmonic resonances and between the resonances the responsivity can be still relatively high. 
Hence, the non-resonant response can  also be useful.

\section{Comments}

Above we assumed that $J_{SD} \propto \sigma \propto \mu$.   Theoretical studies show
that the doped  GC($\mu_D \gg T_0$)  conductivity $\sigma$ can exhibit different dependences on $\mu$~\cite{36,37,38}. In particular,
it can vary from $\sigma$ virtually independent of $\mu$ if the short-range scattering of electrons is dominant 
 to $\sigma \propto \mu^2$ in the case of the long-range scattering (for example, on charged clusters)~\cite{38}.
In the first situation $\nu \propto p$, where $p$ is the electron momentum. In the second
case, $\nu \propto p^{-1}$. 
  In this regard, our model corresponds to an intermediate
 $\sigma$ vs $\mu$ relation (see, for example,~Ref.~37), in which 
 the momentum dependence of $\nu$ is disregarded. This  provides $\sigma \propto \sqrt{\Sigma}\propto \sqrt{V_G}$ ($V_G$ is the voltage swing between the GC and the gate). The latter qualitatively agrees with the experimental data~\cite{1}.  
In such a case, setting $\nu =$~constant (see, for example,Ref.~38), we obtain the relation

\begin{eqnarray}\label{eq27}
\sigma = \frac{e^2T}{\pi\hbar^2\nu}
\int_0^{\infty}d\xi\xi\frac{d}{d\xi}\biggl[-\frac{1}{\exp(\xi - \mu/T) + 1}\biggr]\nonumber\\ = \frac{e^2}{\pi\hbar^2\nu}[\mu + T\ln(1+e^{-\mu/T}) \simeq \frac{e^2\mu}{\pi\hbar^2\nu},
\end{eqnarray}
which was used above.

Considering that 
 the resonant voltage responsivity, $R_{\omega}^{V,GG}$,
 of GC/b-AsP FETs with the biased gate can be estimated as~\cite{16}

\begin{eqnarray}\label{eq28}
R_{\omega}^{V,GG} \sim \frac{16\pi^2}{137}\frac{\hbar}{eT_0}\frac{\tau_{\varepsilon}}{\tau_{\bot}}\frac{\Delta_M}{T_0}\exp\biggl(-\frac{\Delta_M}{T_0}\biggr), 
\end{eqnarray}
for the ratio of the voltage responsivities   we obtain

\begin{eqnarray}\label{eq29}
\frac{R_{\Omega}^V}{R_{\Omega}^{V,GG}} \simeq \frac{\varkappa}{2}
\biggl(\frac{L_{\varepsilon}}{L}\biggr)^2\biggl(\frac{\tau_{\varepsilon}}{\tau_{\bot}}\biggr)\biggl(\frac{T_0}{\mu_D}\biggr)\exp\biggl(\frac{\Delta_M}{T_0}\biggr)
\biggl(\frac{eV_{SD}}{\mu_D}\biggr).\,
\end{eqnarray}
For the typical parameters used above and $V_{SD} \sim (10-25)$~mV,
the latter ratio is about  unity, although it increases with further (linearly) increase in $V_{SD}$. The latter might be limited by the lattice heat removal via the substrate and the contacts. Setting $\nu = (1- 2)$~ps$^{-1}$,
for the thermal power we obtain 
$P_{Th} \sim (0.3  - 0.7)$~mW  at $V_{SD} =  0.2$~V
and 
$P_{Th} \sim (41  - 47)$~mW  at $V_{SD} =  1.6$~V.

As follows from the obtained results, both the current and voltage responsivities are proportional to the source-drain bias voltage $V_{SD}$.
The dark current is also proportional to $V_{SD}$. This implies that the noise-equivalent power (NEP) and the dark current-limited detectivity of the detectors under consideration vary with increasing source-drain voltage as NEP$\propto 1/\sqrt{V_{SD}}$ and  $D_{\Omega}^* \propto \sqrt{V_{SD}}$, respectively.

For the GC/b-As and GC/b-P detectors with the above parameters 
%and $2LH = 4\times10^{-8}~\mu$m$^2$  
at $V_{SD} = 1.6$~V,
we obtain  NEP$ \simeq 2.2$~pW/Hz$^{1/2}$ and
NEP$\simeq 2.5$~pW/Hz$^{1/2}$
(for the GC/b-As and GC/b-P FETs, respectively), which appears to be promising (compare with other THz bolometers~\cite{9}).
If $\sqrt{2LH} = 2~\mu$m, the latter corresponds to $D_{\Omega}^* \simeq 9\times 10^7$~cm Hz$^{1/2}$/W $\simeq 8\times 10^7$
~cm Hz$^{1/2}$/W, 
which is about of or exceeding the detectivity  of other uncooled THz bolometers
(see, for example,~Ref.~39). However, one needs to note that
NEP increases and $D_{\Omega}^*$ decreases with increasing $\nu$.

Since the operation speed of  the  detectors under  consideration
is determined by  the characteristic times of the electron cooling, $t_{\theta}\lesssim \tau_{\varepsilon}/(1+\theta)$, associated with the energy relaxation
on phonons and the 
heat transfer over the BL, and the gate recharging time.
The latter is estimated as $t_{rc} \sim \tau_{\bot}(2T_0\mu_0/\mu_D^2)\exp(\Delta_M/T_0)$. The comparison of these characteristic times yields

\begin{eqnarray}\label{eq30}
 \frac{t_{\theta}}{t_{rc}} \simeq \frac{\tau_{\varepsilon}}{\tau_{\bot}(1+\theta)}
\biggl(\frac{\mu_D^2}{2T_0\mu_0}\biggr)
 \exp\biggl(-\frac{\Delta_M}{T_0}\biggr).
\end{eqnarray}
For the device structural parameters assumed above  $T_0 = 25$, we obtain $t_{\theta}/t_{rc} \sim 0.5$. This implies that the GC/b-AsP FET bolometers response time is about 
$t_r \sim t_{\theta} +t_{rc} \sim 20 $~ps.

The values  of the collision frequencies used in the above calculations
can be expressed via the electron mobility $M$. Using the relation
$M =e v_W^2/\mu_D\nu$, where $m =\mu/v_W^2$ is the so-called fictitious electron mass in GCs, for $\mu_D = 140$~meV and $\nu =(1 -4)$~ps$^{-1}$, we obtain the range
$M \simeq (1.78 - 7.14)\times 10^4$~cm/V$\cdot$s (compare, for example, with ~Refs.~40,  and 41). According to the estimates~\cite{41}, the room temperature mobility
in the GCs on h-BN at the electron density corresponding to the above the Fermi energy can be about $m \simeq 10^5$cm/V$\cdot$s. The latter corresponds to $\nu \simeq 0.714 $~ps$^{-1}$. 
The quality of the interface between the GC (placed atop of h-BN) and the b-P BL can limit
the values of $M$ and $\nu$. The pertinent room temperature electron mobility
obtained experimentally several years ago is equal to 
$M\simeq (7-8)\times 10^3$~cm/Vs~\cite{42} ($M$ exceeds 
$10^4$~cm/Vs
at $T_0 \leq 200$~K). This corresponds to not too small  $\nu$. 
One can expect that the contemporary technology is able to
provide the GC/b-P interface with sufficiently small $\nu$,
at which the parameter ($(\pi\nu/2\Omega)^2 < 1$, so that the plasmonic resonances are pronounced.
A substantial reinforcement of the plasmonic resonances in the GC-FET detectors can be realized in the case of the composite gate BL,
which includes a relatively narrow b-P BL (and the MG) and
the h-BN (or WSe$_2$~\cite{43}) sections between the b-P section and the source in drain.
In such a GC-FET detector, the sharpness of the  plasmonic
resonances might be determined by the GC  main part (encapsulated by h-BN and providing small electron collision frequency), by the
thermionic current flows via the narrow b-P window.

Similar THz detection properties can be expected in the GC/b-AsP FET devices with
the floating isolated doped graphene gate (GG). The main distinction between
the detectors  with the MG and the detectors with the GG is the different plasmonic response in the latter  because of the GG influence on  the plasmonic
oscillations in the double-graphene structures~\cite{44,45,46}.  
Another option is to use the MG consisting of an array of metallic islands (MIs) or quantum dots (QDs). In such a case, each  MI/QD has its own floating potential determined by the electron exchange between the MG/QD and the GC (compare with the devices analyzed in Ref.~9). Due to this,
the potential distribution along the GC and the effect of the floating MG on the source-drain current
can be markedly different from
that considered above.
However, the consideration of the detectors in question
 requires a proper modification of the device model and, therefore, a separate treatment.

\section{Conclusions}

We estimated the room temperature characteristics of the proposed GC/b-AsP FETs with the floating MG operating in the THz frequency range at room temperature. We showed that these detectors can exhibit  high values of  responsivity at the  plasmonic resonances ($ \gtrsim  10^3$~V/W) and rather short response times ($\sim 20$~ps).

\section*{Author contributions}
All authors contributed equally to this work.

\section*{Acknowledgments}

The Japan Society for Promotion of Science (KAKENHI
Grants $\#$ 21H04546 and $\#$ 20K20349), Japan;
RIEC Nation-Wide Collaborative Research Project $\#$
R04/A10; the US Office of Scientific, Research Contract
N00001435, (Project Monitor Dr. Ken Goretta).

\section*{Conflict of Interest}

The authors declare no conflict of interest
\newpage
\section*{Data availability}
All data that support the findings of this study
are available within the article.


\begin{thebibliography}{1}
\section*{References}
\bibitem{1} A. H. Castro Neto, F. Guinea, N. M. Peres, K. S.
Novoselov, and A. K. Geim, \lq\lq The electronic properties
of graphene,\rq\rq\, Rev. Mod. Phys. {\bf 81}, 109 (2009).

\bibitem{2}
A. Zolnierski, W. Cetera, D. Jaruga, J. Grzegorek,
and  G. Sowula, \lq\lq Graphene and its applications. Study on the development trends in
research and on the implementation potential using Big Data and
information refining methods,\rq\rq\,
J. Nanotechnol. Res. {\bf 4}, 117 (2022). 


\bibitem{3}
N. M. R. Peres,
\lq\lq The transport properties of graphene,\rq\rq\,
J. Phys. Cond. Mat. {\bf 21}, 323201 (2009).


\bibitem{4}
R. W. Keyes, 
\lq\lq
The electrical properties of black phosphorus,\rq\rq Phys. Rev. {\bf 92}, 580 (1953).

\bibitem{5}
H Asahina and A Morita,
\lq\lq
Band structure and optical properties of black
phosphorus,\rq\rq\,
 J. Phys. C: Solid State Phys. {\bf 17} 1839 (1984).

\bibitem{6}
Xi Ling, H. Wang, S. Huang, F. Xia, and M. S. Dresselhaus,
\lq\lq
The renaissance of black phosphorus,\rq\rq\,
PNAS {\bf 122}, 4523 (2015).


 
\bibitem{7}
Y. Cai, G. Zhang, and Y.-W. Zhang, 
\lq\lq
Layer-dependent band alignment and work function of few-layer phosphorene,\rq\rq\,
 Sci. Rep. {\bf 4}, 6677 (2015). 



\bibitem{8}
 F. Xia, H. Wang, and Y. Jia, 
 \lq\lq Rediscovering black phosphorous as an anisotropic layered material for optoelectronics and electronics,
 \rq\rq\,
  Nat. Commun. {\bf 5}, 4458 (2014).   
 
\bibitem{9}
A. Rogalski, M. Kopytko, and P. Martyniuk,
\lq\lq Two-dimensional infrared and terahertz detectors: Outlook and status,\rq\rq\, Appl. Phys. Rev. {\bf 6}, 021316 (2019); 


\bibitem{10} 
V. Ryzhii, M. Ryzhii, D. Svintsov, V. Leiman, P. P. Maltsev, D. S. Ponomarev, V. Mitin, M. S. Shur, and T. Otsuji,
\lq\lq
Real-space-transfer mechanism of negative differential conductivity in gated graphene-phosphorene
hybrid structures: Phenomenological heating model,\rq\rq\,
J. Appl. Phys. {\bf 124}, 114501 (2018).


\bibitem{11} 
A. Parappurath, S. Mitra, G. Singh, N. K. Gill,
T. Ahmed, T. P. Sai, K. Watanabe, T. Taniguchi, and
A. Ghosh,
\lq\lq
Interlayer charge transfer and photodetection efficiency
of graphene-transition-metal-dichalcogenide heterostructures,\rq\rq\,
Phys. Rev. Appl. {\bf 17}, 064062 (2022). %% BUT single GL!


\bibitem{12} Z. Guo, H. Zhang, S. Lu, Z. Wang, S. Tang, J. Shao, Z. Sun, H. Xie, H. Wang, X.-F. Yu, and P. K. Chu, 
\lq\lq
From black phosphorus to phosphorene: Basic solvent exfoliation, evolution of Raman scattering, and applications to ultrafast photonics,\rq\rq\,
 Adv. Funct. Mater. {\bf 25}, 6996 (2015).  

\bibitem{13}
 M. Long, A. Gao, P. Wang, H. Xia, C. Ott, C. Pan, Y. Fu, E. Liu, X. Chen, W. Lu, T. Nilges, J. Xu, X. Wang, W. Hu, and F. Miao, 
\lq\lq  Room temperature high-detectivity mid-infrared photodetectors based on black arsenic phosphorus,\rq\rq\,	
 Sci. Adv. {\bf 3}, e1700589 (2017). 



\bibitem{14} 
V. Ryzhii, T. Otsuji, M. Ryzhii, D. S. Ponomarev, V. E. Karasik,
V. G. Leiman, V. Mitin, and M. S. Shur,
\lq\lq
Electrical modulation of terahertz radiation
using graphene-phosphorene
heterostructures,\rq\rq\,
Semicond. Sci. Technol. {\bf 33}, 124010 (2018).







\bibitem{15}
V. Ryzhii, M. Ryzhii V. G. Leiman,
V. E. Karasik,  V. Mitin,
M. S. Shur, and T. Otsuji,
\lq\lq
Far-infrared photodetectors based on
nanoribbon heterostructures with black-AsP
barrier layers,\rq\rq\,
Opt. Exp. {\bf 28}, 2480 (2020).



\bibitem{16}
V. Ryzhii, C. Tang, T. Otsuji, M. Ryzhii, V. Mitin, and
M. S. Shur, \lq\lq Resonant plasmonic detection of terahertz
radiation in field-effect transistors with the graphene
channel and the 
black-AsP gate layers,
\rq\rq\,(submitted), arXiv: 2304.11635v1.

\bibitem{17} 
U. Sassi, R. Parret, S. Nanot, M. Bruna S. Borini, D. De Fazio,
Z. Zhao,
E. Lidorikis, F. H. L. Koppens,
A. C. Ferrari, and   A. Colli, \lq\lq Graphene-based mid-infrared room-temperature
pyroelectric bolometers with ultrahigh temperature coefficient of resistance,\rq\rq\, Nat. Commun.
{\bf 8}, 14311 (2017). 
%https://doi.org/10.1038/ncomms14311


 \bibitem{18}
V. Ryzhii, A. Satou, and T. Otsuji,
\lq\lq Plasma waves in two-dimensional electron-hole system in gated graphene heterostructures,\rq\rq\,
J. Appl. Phys. {\bf 101}, 024509 (2007).

 \bibitem{19}
A. N. Grigorenko, M. Polini, and K. S. Novoselov, \lq\lq Graphene plasmonics,\rq\rq\,
 Nat. Photonics {\bf 6}, 749 (2012).
 
  \bibitem{20}  
A.V. Muraviev, S. L. Rumyantsev, G. Liu, A. A. Balandin, W. Knap, and M. S. Shur, \lq\lq  
 Plasmonic and bolometric terahertz detection by graphene field-effect transistor,\rq\rq\,
  Appl. Phys. Lett. {\bf 103}, 181114 (2013).
  %https://doi.org/10.1063/1.4826139


\bibitem{21} 
V. Ryzhii, T. Otsuji, M. Ryzhii, V. Mitin, and M. S. Shur,
\lq\lq Resonant plasmonic terahertz detection in vertical graphene-base hot-electron
transistors,\rq\rq\,
J. Appl. Phys. 118, 204501 (2015).


\bibitem{22} 
 V. Ryzhii, T. Otsuji, and M. S. Shur,
 \lq\lq Graphene based plasma-wave devices for terahertz applications,\rq\rq\,
 Appl. Phys. Lett. {\bf 116}, 140501 (2019).


\bibitem{23}
S. Luryi, \lq\lq Quantum capacitance devices,\rq\rq\,
Appl. Phys. Lett. {\bf 52}, 501 (1988).



\bibitem{24}
S. Dr\"{o}scher, P. Roulleau, F. Molitor, P. Studerus, C. Stampfer, K. Ensslin, and T. Ihn,
\lq\lq Quantum capacitance and density of states of graphene,\rq\rq\,
Appl. Phys. Lett. 96, 152104 (2010). %10.1063/1.3391670\\









\bibitem{25} 
Z. Tong, A. Pecchia, C. Yam, T. Dumitrică, and  T. Frauenheim,
\lq\lq Ultrahigh electron thermal conductivity in T-Graphene, Biphenylene, and Net-Graphene,\rq\rq\,
Adv. Energy Mater. {\bf 12}, 2200657 (2022).

\bibitem{26}
T. Y. Kim, C.-H. Park, and N. Marzari,\lq\lq
The electronic thermal conductivity of graphene,\rq\rq\,
Nano Lett. {\bf 16}, 2439-2443 (2016). 






\bibitem{27} 
J. H. Strait, H. Wang, S. Shivaraman, V. Shields, M. Spencer, and F. Rana,
\lq\lq
Very slow cooling dynamics of photoexcited carriers in graphene observed by optical-pump terahertz-probe spectroscopy,\rq\rq\,
Nano Lett. {\bf 11}, 4902 (2011).


\bibitem{28}
V. Ryzhii, M. Ryzhii, V. Mitin, A. Satou, and T. Otsuji,
\lq\lq
 Effect
of heating and cooling of photogenerated electron-hole plasma
in optically pumped graphene on population inversion,\rq\rq\, 
Jpn. J. Appl. Phys. {\bf 50}, 094001 (2011).

\bibitem{29} 
V. Ryzhii, T. Otsuji, M. Ryzhii, M. Ryzhii, N. Ryabova, S. O. Yurchenko, V. Mitin, and M. S. Shur,
\lq\lq Graphene terahertz uncooled bolometers,\rq\rq\,
J. Phys. D: Appl. Phys {\bf 46}, 065102 (2013).

\bibitem{30} 
V. Ryzhii, A. Satou, T. Otsuji, M. Ryzhii, V. Mitin, and M. S. Shur,
\lq\lq Graphene vertical hot-electron terahertz detectors,\rq\rq\,
J. Appl. Phys. {\bf 116}, 114504 (2014).

\bibitem{31} 
V. Ryzhii, T. Otsuji, M. Ryzhii, A. A. Dubinov, V. Ya. Aleshkin, V. E. Karasik, and M. S. Shur,
\lq\lq
Negative terahertz conductivity and amplification of surface plasmons in graphene-black
phosphorus injection laser heterostructures,
\rq\rq\,
Phys. Rev. B {\bf 100}, 115436 (2019). %%% T


\bibitem{32}  J. F. Rodriguez-Nieva, M. S. Dresselhaus, and L. S. Levitov,
\lq\lq Thermionic emission and negative dI/dV in photoactive
graphene heterostructures,\rq\rq\, 
Nano Lett., {\bf 15} 145 (2015).
%%%%%Hot Carriers  mainly due to radiation interband heating - NDR between GLs

\bibitem{33}
V. Ryzhii, T. Otsuji, M. Ryzhii, V. Ya. Aleshkin, A. A. Dubinov,  V. Mitin, and M. S. Shur,
\lq\lq Graphene electron transport in van der Waals heterostructures with graphene layers,\rq\rq\,
J.  Appl. Phys. {\bf 117}, 154504 (2015).





\bibitem{34}
 M. Uda, A. Nakamura, T. Yamamoto, and Y. Fujirnoto,
\lq\lq Work function of polycrystalline Ag, Au, and A1,\rq\rq\,
J. Electron Spectros. Relat. Phenomena {\bf 88-91}, 643 (1998).





\bibitem{35}
B. Liu, M. K{\"o}pf, A. N. Abbas, X. Wang, Q. Guo, et al.,
% Yichen Jia ,
%Fengnian Xia , Richard Weihrich , Frederik Bachhuber , Florian Pielnhofer , Han Wang ,
%Rohan Dhall , Stephen B. Cronin , Mingyuan Ge , Xin Fang , Tom Nilges ,
%and Chongwu Zhou,
\lq\lq
Black Arsenic-Phosphorus: Layered anisotropic infrared
semiconductors with highly tunable compositions
and properties,\rq\rq\,
Adv. Mater. {\bf 27}, 4423 (2015).




\bibitem{36}
F. T. Vasko and V. Ryzhii,
\lq\lq Voltage and temperature dependences of conductivity in gated graphene heterostructures,\rq\rq\,
Phys. Rev. B {\bf 76}, 233404 (2007).

\bibitem{37}
L. A. Falkovsky,
\lq\lq Optical properties of graphene and IV-VI semiconductors,\rq\rq\,
Physics-Uspekhi {\bf 51}, 887 (2008).


\bibitem{38}
V. Ryzhii, D. S. Ponomarev, M. Ryzhii, V. Mitin, M.S. Shur, and T. Otsuji,
\lq\lq Negative and positive terahertz and infrared
photoconductivity in uncooled graphene,\rq\rq\,
Opt. Mat. Exp. {\bf 9}, 585 (2019).


%\bibitem{38}
%D. Svintsov, V. Vyurkov, S. Yurchenko, T. Otsuji, and V. Ryzhii,
%\lq\lq Hydrodynamic model of electron-hole plasma in graphene,\rq\rq\,
%J. Appl. Phys. {\bf 111}, 083715 (2012).

\bibitem{39}
 A. Rogalski,
\lq\lq Semiconductor detectors and focal plane arrays for far-infrared imaging,\rq\rq\,
Opto-Electron. Rev. {\bf 21}, 406 (2013).



%H. Yamashita, K. Tsunoda, H. Nishino, and S. Sato,
%\lq\lq Signal-to-noise characteristics of graphene photodetectors based on photothermic effect,\rq\rq\,
%J. Appl. Phys. {\bf 129}, 173101 (2021).


\bibitem{40}
A. S. Mayorov, R. V. Gorbachev, S. V. Morozov,
L. Britnell, R. Jalil, L. A. Ponomarenko, P. Blake,
K. S. Novoselov, K. Watanabe, T. Taniguchi, and
A. K. Geim, \lq\lq Micrometer-scale ballistic transport in encapsulated
graphene at room temperature,\rq\rq\,Nano Lett.,
vol. 11, 2396 (2011).



\bibitem{41}
 H. Hirai, H. Tsuchia, Y. Kamakura, N. Mori, and M. Ogawa, \lq\lq Electron mobility calculation for graphene on substrates,\rq\rq\, J.  Appl. Phys.
{\bf 116}, 083703 (2014). %%%; https://doi.org/10.1063/1.4893650
 
 
 
 
\bibitem{42}
Y. Liu, I. Yudhistira, M. Yang, E. Laksono, Y. Z. Luo,
J. Chen, J. Lu, Y. P. Feng, S. Adam, and K. P. Loh, \lq\lq Mediated colossal magnetoresistance in
graphene/black phosphorus heterostructures,\rq\rq Nano Lett.
{\bf 18}, 3377 (2018).



\bibitem{43}
L. Banszerus, T. Sohier, A. Epping, F. Winkler,
F. Libisch, F. Haupt, K. Watanabe, T.
Taniguchi, K. M¨uller-Caspary, N. Marzari, F. Mauri,
B. Beschoten, and C. Stampfer,“ Extraordinary high
room-temperature carrier mobility in graphene-WSe2
heterostructures,”arXiv:1909.09523v1/

%%ZZZZZZZ

\bibitem{44}
V. Ryzhii, T. Otsuji, M. Ryzhii, and M. S. Shur,
\lq\lq
Double graphene-layer plasma resonances
terahertz detector,\rq\rq\,
J. Phys. D: Appl. Phys. {\bf 45}, 302001 (2012). %%%%%%THz Using Tunneling J_tunn \propt to V^3



\bibitem{45} 
V. Ryzhii, A. Satou, T. Otsuji, M. Ryzhii, V. Mitin, and M. S. Shur,
\lq\lq 
Dynamic effects in double graphene-layer
structures with inter-layer
resonant-tunneling negative conductivity,\rq\rq\,
J. Phys. D: Appl. Phys. {\bf 46}, 315107 (2013). %%%% The same as above, but RT. 


\bibitem{46} 
V. Ryzhii, M. Ryzhii, V. Mitin, M. S. Shur, A. Satou, and T. Otsuji,
\lq\lq
Terahertz photomixing using plasma resonances in double-graphene layer
structures,\rq\rq\,
J. Appl. Phys. {\bf 113}, 174506 (2013).


 \end{thebibliography}
\end{document}